\documentclass[aps,prd,floatfix,nofootinbib,superscriptaddress,showpacs,twocolumn]{revtex4}
\usepackage{exscale}
\usepackage[intlimits]{amsmath}
\usepackage{amsfonts}
\usepackage{amssymb,amscd}                
\usepackage{array}
\usepackage{wrapfig}
\usepackage{bbm}
\usepackage{multirow}
\usepackage[x11names]{xcolor}
\usepackage{graphicx}
\usepackage[T1]{fontenc}
\usepackage{times}
\usepackage{subfigure}

\newcommand{\sh}[1]{#1\hskip-7pt \diagup}
\newcommand{\Sh}[1]{#1\hskip-10pt \diagup}
\newcommand{\tr}{\textrm{tr}}
\newcommand{\F}{\mathcal{F}}

\newcommand{\Eq}[1]{Eq.~(\ref{#1})}

\newcommand{\lqcd}{\Lambda_{\mathrm{QCD}}}
\newcommand{\lqcdsq}{\Lambda^2_{\mathrm{QCD}}}

\definecolor{TGcolor}{rgb}{0.6,0.6,0.0}

\begin{document}

\title{Hadronic light-by-light scattering in the muon $g-2$: a Dyson-Schwinger equation approach}

\author{Tobias Goecke}
\affiliation{Institute for Nuclear Physics, 
 Darmstadt University of Technology, 
 Schlossgartenstra{\ss}e 9, 64289 Darmstadt, Germany}
\author{Christian S. Fischer}
\affiliation{Institut f\"ur Theoretische Physik, 
 Universit\"at Giessen, 35392 Giessen, Germany}
\affiliation{Gesellschaft f\"ur Schwerionenforschung mbH, 
  Planckstr. 1  D-64291 Darmstadt, Germany.}
\author{Richard Williams}
\affiliation{Institute for Nuclear Physics, 
 Darmstadt University of Technology, 
 Schlossgartenstra{\ss}e 9, 64289 Darmstadt, Germany}

\begin{abstract}
We determine the hadronic light-by-light scattering contribution 
to the anomalous magnetic moment of the muon using the framework of 
Dyson-Schwinger and Bethe-Salpeter equations of QCD. Our result for 
the pseudoscalar ($\pi^0, \eta, \eta'$) meson exchange diagram
is commensurate with previous calculations. In our calculation of the
quark loop contribution we improve upon previous approaches by 
explicitly implementing constraints due to gauge invariance. The
impact of transverse contributions, presumably dominated by vector 
meson poles, are only estimated at this stage. As a consequence, our value 
$a_\mu^{\textrm{LBL;quarkloop}} = (136 \pm 59)\times 10^{-11}$
is significantly larger. Taken at face value, this then leads to a 
revised estimate of the total $a_\mu=116\,591\,891.0(105.0)\times 10^{-11}$.
\end{abstract}
\pacs{12.38.Lg, 13.40.Em, 13.40.Gp, 14.60.Ef}
\maketitle

\section{Introduction\label{sec:intro}}
One of the most impressive successes of the standard model of particle physics
is the determination of the anomalous magnetic moment of the electron. This
quantity is determined both experimentally and theoretically to such a degree of 
precision that the underlying physical description is vindicated. However, 
when it comes to the question of new physics, the anomalous magnetic moment
of the muon is an even more interesting quantity, 
see e.g. \cite{Jegerlehner:2009ry,Jegerlehner:2008zza,Stockinger:2006zn} 
for reviews. This is due to the large mass of the muon as compared to the
electron, which leads to an enhanced sensitivity to physics in and beyond the
standard model. Experimental
efforts at Brookhaven and theoretical efforts of the past ten years have
pinned $a_\mu$ down to the $10^{-11}$ level, leading to significant deviations
between theory \cite{Jegerlehner:2009ry} and experiment \cite{Bennett:2006fi}:
      \begin{align}
		\label{eqn:amuexperiment}
            \mbox{Experiment:} \,\,\,\,
			&116\,592\,089.0(63.0)\times 10^{-11}\;\;, 
		\\
		\label{eqn:amutheoretical}
            \mbox{\phantom{wwu}} \mbox{Theory:} \,\,\,\,
			&116\,591\,790.0(64.6)\times 10^{-11}\;\;.
      \end{align}
Whilst the theoretical and experimental values are determined to comparable
errors, the central values give rise to a discrepancy at the $3.3\,\sigma$
confidence level. This difference has been present for a number of years and
can be interpreted as a signal for the existence of physics beyond the
standard model. However, to clearly distinguish between New Physics
and possible shortcomings in the SM calculations the uncertainties 
present in both experimental and theoretical values of $a_\mu$ 
need to be further reduced.

\begin{figure}[t!]
			\subfigure[][]{\label{fig:hadroniclo}\includegraphics[width=0.40\columnwidth]{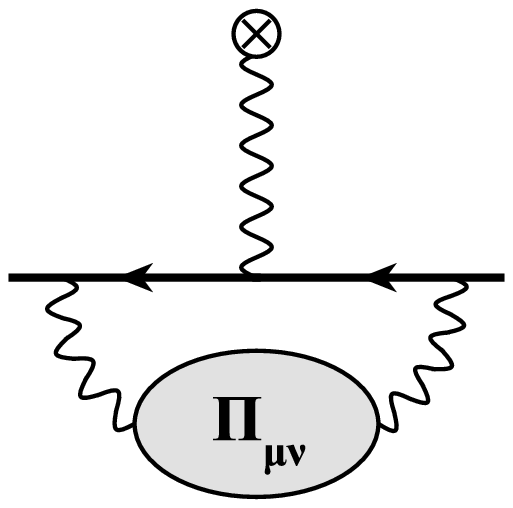}}
	\hspace{0.08\columnwidth}
	\subfigure[][]{\label{fig:hadroniclbl}\includegraphics[width=0.40\columnwidth]{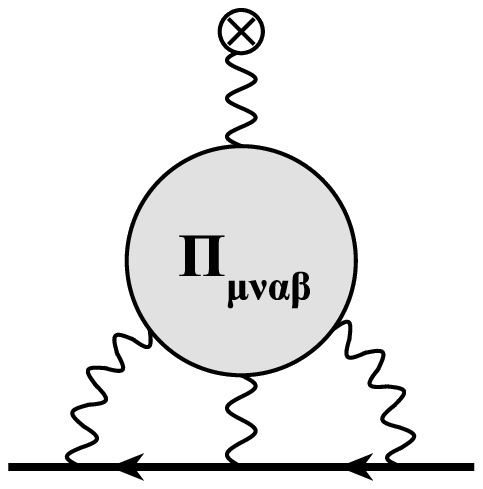}}
      \caption{The two classifications of corrections to the photon-muon
	vertex function: (a) hadronic vacuum polarisation contribution to $a_\mu$. The vertex is 
               dressed by the vacuum polarisation tensor $\Pi_{\mu\nu}$;
		   (b) the hadronic light-by-light scattering contribution to
		   $a_\mu$.}
\end{figure}

The greatest uncertainties in the theoretical determination of $a_\mu$
are encountered in the hadronic contributions, i.e.
those terms which involve QCD beyond perturbation theory. The 
most prominent of these is given by the vacuum polarisation tensor 
dressing of the QED vertex, see Fig.~\ref{fig:hadroniclo}. Fortunately it can be
related to experimental data of $e^+ e^-$-annihilation and $\tau$-decay via 
dispersion relations and the optical theorem, thus resulting in a precise 
determination with systematically improvable errors \cite{Jegerlehner:2009ry}.
Considered individually, its (leading and subleading order) contribution to 
the anomalous magnetic moment of
the muon is~\cite{Jegerlehner:2008zza}
      \begin{align}\label{eqn:hadroniclo}
            [6\,903.0(52.6) - 100.3(1.1)]\times 10^{-11}\;\;.
      \end{align}
Although currently these uncertainties dominate the error
of the theoretical result in \Eq{eqn:amutheoretical} it is foreseeable that 
future experiments reduce this error below that of another, more problematic
source. This is the hadronic light-by-light (LBL) scattering 
diagram, shown in Fig.~\ref{fig:hadroniclbl}. This contribution cannot be directly 
related to experiment and must hence be calculated entirely through theory.
\begin{table}[b]
	\centering
	\renewcommand{\tabcolsep}{1.0pc} 
	\renewcommand{\arraystretch}{1.1} 

	\begin{tabular}{@{}ll}
            { Ref.} & $a_\mu^{\mathrm{LBL}}$\\
            \hline
            \cite{Nyffeler:2009tw}                                 
                  & $116(40)\times10^{-11}$    \\
            \cite{Prades:2009tw}                                   
                  & $105(26)\times10^{-11}$                         \\
            \cite{Bijnens:2007pz}                                  
                  & $110(40)\times10^{-11}$   \\
            \cite{Hayakawa:1995ps,Hayakawa:1996ki,Hayakawa:1997rq} 
                  & $89(15)\,\,\,\times10^{-11}$
      \end{tabular}
\caption{Recent calculations of the hadronic light-by-light scattering contribution to the anomalous magnetic moment of the muon.\label{tab:recent}}
\end{table}
The central object in such a calculation is the photon four-point
function. It receives important contributions from
the small momentum region below $2$~GeV, where perturbative QCD breaks down and 
non-perturbative methods are imperative. Recent determinations of $a_{\mu}^{\mathrm{LBL}}$ 
are provided in Table~\ref{tab:recent}. Although the magnitude of the LBL 
contribution is much smaller than the one from vacuum polarization it is 
significant because its error is of a comparable size. Taken together, with 
the errors added in quadrature, the hadronic contributions constitute the
largest uncertainty in the standard model determination of the anomalous
magnetic moment of the muon. 

The theoretical approaches to determine the LBL contribution are centered around
two main ideas. One is chiral symmetry, its breaking pattern and the associated
low energy effective descriptions of QCD \cite{Ecker:1989yg}; the other is the large-$N_c$ expansion
of the four-photon function and the associated ordering of diagrams. These ideas
have been put together in \cite{deRafael:1993za} and 
led to various refined calculations of LBL within the frameworks of large-$N_c$ 
and vector meson dominance~\cite{Knecht:2001qf,Melnikov:2003xd,Nyffeler:2009tw,Nyffeler:2010rd}, 
the extended Nambu-Jona-Lasinio model (ENJL) \cite{Bijnens:1995cc,Bijnens:2007pz}, 
the (very similar) hidden local symmetry model \cite{Hayakawa:1995ps,Hayakawa:1996ki,Hayakawa:1997rq}, 
or a non-local chiral quark model \cite{Dorokhov:2004ze,Dorokhov:2008pw}, 
see also \cite{Prades:2009tw} for a summary. Although in terms of diagrams 
individual contribution are of varying size in these approaches, their sum leads to 
consistent results as can be inferred from Table~\ref{tab:recent}.  In all these calculations the (pseudoscalar) meson
exchange contributes the most and the meson loop has been found to be small. A
possible explanation of the latter is given in \cite{Melnikov:2003xd}.
As a result, we quote the recent value for LBL 
$a_\mu^{\mathrm{LBL}} = 105(26)\times10^{-11}$ proposed in Ref.~\cite{Prades:2009tw}, which
also agrees with the one in \cite{Nyffeler:2009tw}.

One of the most important goals for these and future calculations is the reduction of the
model dependence and subsequently of the systematic error involved in these calculations. 
Since LBL is non-perturbative in nature, all estimates in Table~\ref{tab:recent} are
plagued by systematic model dependencies. It is therefore desirable to also 
explore other calculational tools which have the potential to go beyond these limitations. 
Certainly, lattice gauge theory is one such method. However, due to the multi-scale
nature of the problem no reliable estimates for LBL have been extracted on the lattice so far.
This multi-scale nature also makes EFT methods less desirable as it
proves more difficult to impose suitable matching conditions.

Another non-perturbative method, well suited to accommodate for largely 
different scales is the framework of Dyson-Schwinger and Bethe-Salpeter 
equations~\cite{Alkofer:2000wg,Maris:2003vk,Fischer:2006ub,Roberts:2007ji}. 
In the past years this approach has been used to study fundamental properties 
of QCD such as confinement and dynamical chiral symmetry breaking. On the other 
hand the approach served as a tool for hadron physics. In this work we
expand upon this and apply the formalism to a calculation of the LBL 
contribution to the muon anomalous magnetic moment. To this end we separate 
different contributions to the light-by-light four-point function according to
their topology of gluon exchange and their status with respect to the large-$N_c$
expansion. Diagrammatically this translates to considering resummations of 
planar diagrams involving gluon exchange. In this scheme we then
determine the dressed quark-loop diagram and an approximation in terms of 
pseudoscalar meson ($\pi^0, \eta, \eta'$) exchange contributions. In principle,
the off-shell meson amplitudes involved in these diagrams could be calculated from
inhomogeneous Bethe-Salpeter equations. Here, due to numerical complexity, in this
work we resort to a commonly used ansatz that extrapolates on-shell wave functions.
Our results are then compared with the ones of previous approaches. First results of our
analysis have been published in Ref.~\cite{Fischer:2010iz}. Here we discuss our
method in much more detail and present new and more elaborate results for the
quark-loop diagram.

The paper is organized thus: in section \ref{sec:lbl} we recall the definition of the 
light-by-light scattering amplitude and focus upon the pseudoscalar pole contributions; 
in section~\ref{sec:model} we introduce our Dyson-Schwinger approach and discuss the 
necessary truncation schemes; in section~\ref{sec:results} we present and 
discuss our results. We conclude in section~\ref{sec:conclusion}.

\section{The LBL scattering amplitude\label{sec:lbl}}
In the hadronic light-by-light scattering contribution, Fig.~\ref{fig:hadroniclbl},
the muon is coupled to an external photon source via the hadronic photon
four-point function $\Pi_{\mu\nu\alpha\beta}$, defined through
      \begin{align}\label{eqn:photon4ptfn}
            \Pi_{\mu\nu\alpha\beta}(q_1,q_2,q_3)&=  \notag\\
            \int_{xyz}\,e^{iq_1\cdot x+iq_2\cdot y+iq_3\cdot z}
              &  \left< j_\mu(0)j_\nu(x)j_\alpha(y)j_\beta(z)\right>\;\;,
      \end{align}
where 
$\int_{xyz}=\int\!d^4x\int\!d^4y\int\!d^4z$
represents integration over four-dimensional space, $q_{1,2,3}$ are the photon momenta
that are connected to the muon line,
and $j_\mu$ is the electromagnetic quark current
      \begin{align}  \label{eqn:quarkcurrent}
            j_\mu &= \frac{2}{3}\bar{u}\gamma_\mu u 
                   - \frac{1}{3}\bar{d}\gamma_\mu d 
                   - \frac{1}{3}\bar{s}\gamma_\mu s
                   + \frac{2}{3}\bar{c}\gamma_\mu c\;\;.
      \end{align}
A detailed discussion of this object can be found in the literature,
see e.g. \cite{Bijnens:1995cc,Melnikov:2003xd}. Instead of working 
directly with the light-by-light scattering diagram
given in Fig.~\ref{fig:hadroniclbl}, it is more convenient to follow
the strategy employed in Ref.~\cite{Aldins:1969jz,Aldins:1970id}.
Here gauge symmetry is exploited to construct quantities that are finite.
Through use of the Ward-Takahashi-identity $k_\mu\Pi_{\mu\nu\alpha\beta} = 0$ 
it follows via differentiation that
	\begin{align}
  		\Pi_{\rho\nu\alpha\beta} =& - k_\mu\frac{\partial}{\partial k_\rho}\Pi_{\mu\nu\alpha\beta} \notag\\
  		=:&-k_\mu \widetilde{\Pi}_{(\rho)\mu\nu\alpha\beta},
  		\label{eqn:DefOfFivePointFunction}
	\end{align}
which serves as definition of the five-point-function $\widetilde{\Pi}_{(\rho)\mu\nu\alpha\beta}$. Here
$k= q_1+q_2+q_3$ is the momentum of the external photon. The virtue of
the derivative is that it lowers the dimensionality of the integral thus
ensuring that integrals employing $\widetilde{\Pi}_{(\rho)\mu\nu\alpha\beta}$ are manifestly
convergent. We define the quantity
	\begin{align}
  		ie\widetilde{\Gamma}_{\rho\mu} = \int_{q_1}\int_{q_2}&
			D_{\epsilon\nu}(q_1)D_{\delta\alpha}(q_2)D_{\gamma\beta}(q_3)\notag\\
           \times&(ie\gamma_\gamma) S(p_1) (ie\gamma_\delta) S(p_2) (ie\gamma_\epsilon)  \notag\\
	   \times&\left[  (ie)^4\widetilde{\Pi}_{(\rho)\mu\nu\alpha\beta}(q_1,q_2,q_3) \right],
	  \label{eqn:GammaRhoMu}
	\end{align}
which is now related to the dressed muon vertex $\Gamma_\mu$ of
Fig.~\ref{fig:hadroniclbl} via
	\begin{align}
 		ie\Gamma_\mu= iek_\rho \widetilde{\Gamma}_{\rho\mu}.
  		\label{eqn:DressedMuonVertex}
	\end{align}
Here $D_{\mu\nu}(q)$ are perturbative photon propagators (we use Feynman
gauge) with momenta $q_i$. The perturbative muon
propagators are given by $S(p)$.  

      \begin{figure*}[t]
            \begin{eqnarray*}
            \begin{array}{c}
            \includegraphics[height=2.2cm]{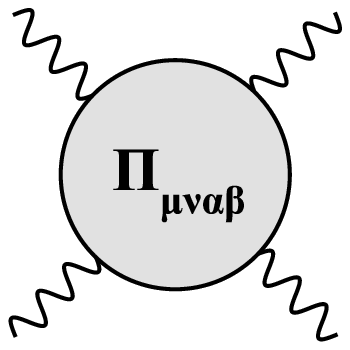}
            \end{array}
            \simeq
            \begin{array}{c}
            \includegraphics[height=2.2cm]{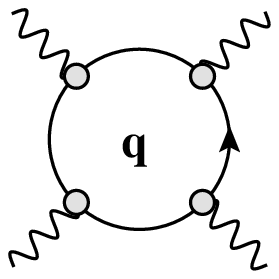}
            \end{array}
            +
            \begin{array}{c}
            \includegraphics[height=2.2cm]{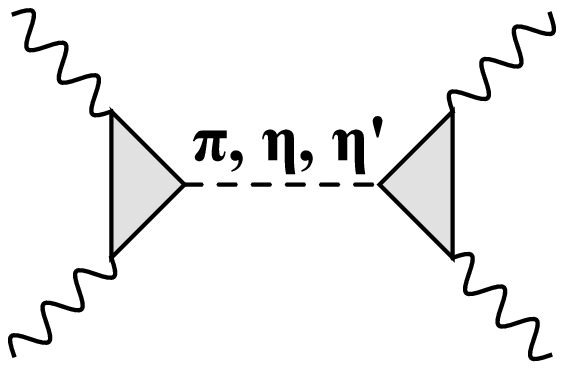}
            \end{array} 
            +          
            \begin{array}{c}
            \includegraphics[height=2.2cm]{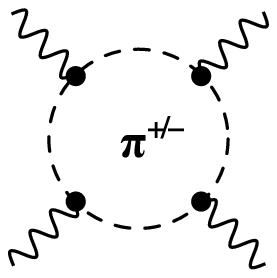}
            \end{array}           
		+\cdots
            \end{eqnarray*}
      \caption{The hadronic light-by-light (LBL) scattering contribution to
               $a_\mu$ and its expansion, using EFT approaches, as a quark loop part (left), 
               leading pseudoscalar meson exchange part (middle) and a
		   leading meson loop part (right). Note that the quarks
		   here may be interpreted differently to those in
		   Fig.~\ref{fig:photon4pt_2}.\label{fig:photon4pt}}
      \end{figure*}
	\begin{figure*}[t!]
            \begin{eqnarray*}
            \begin{array}{c}
            \includegraphics[height=2.2cm]{photon4ptfn}
            \end{array}
            \simeq
            \begin{array}{c}
            \includegraphics[height=2.2cm]{photon4ptfn-qrkloop}
            \end{array} 
            +          
            \begin{array}{c}
            \includegraphics[height=2.2cm]{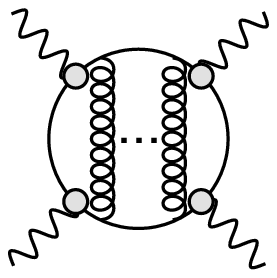}
            \end{array} 
		+
            \begin{array}{c}
            \includegraphics[height=2.2cm]{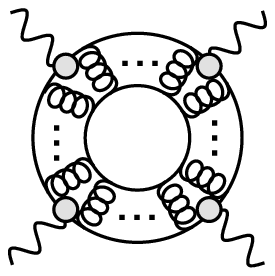}
            \end{array} 
		+ \cdots
            \end{eqnarray*}
      \caption{The hadronic light-by-light (LBL) scattering contribution to
               $a_\mu$ and its expansion, using functional methods, as a quark loop part (left), 
               a ladder exchange part (middle) and a ladder ring 
		   part (right). All propagators and vertices are
		   fully-dressed, with the ellipsis marks indicating that an infinite
		   number of gluons are resummed. \label{fig:photon4pt_2}}
	\end{figure*}

The anomalous magnetic moment can now be obtained by applying the
appropriate projection operator to Eq.~(\ref{eqn:GammaRhoMu})
	\begin{align}
  		a_\mu = \left.\frac{1}{48m_\mu}\tr\left[(i\Sh{P}+m_\mu)[\gamma_\sigma,\gamma_\rho]
  		(i\Sh{P}+m_\mu) \widetilde{\Gamma}_{\sigma\rho}\right]\right|_{k\equiv0},
  		\label{eqn:DefinitionOfAnomaly}
	\end{align}
that we write here in Euclidean convention for later convenience. Using
Eqs.~(\ref{eqn:DefOfFivePointFunction}--\ref{eqn:DefinitionOfAnomaly})
we are able to evaluate the light-by-light scattering contributions for
an arbitrary photon four-point function. What remains now is the
specification of this four-point function within our approach.

\subsection{Expansion using EFT approaches}
As already mentioned in 
the introduction, chiral and large-$N_c$ arguments have been established
to expand the full LBL scattering amplitude into the diagrammatic parts
shown in Fig.~\ref{fig:photon4pt} \cite{Jegerlehner:2009ry}. These diagrams 
belong to different orders with respect to chiral and large-$N_c$ counting. Whereas 
the meson exchange diagrams and the quark-loop diagram are leading in 
large-$N_c$, it is the meson-loop diagram that is leading in the chiral 
counting. Thus \emph{a priori}, one does not know which expansion is to be 
preferred. Therefore it is certainly interesting that all explicit calculations 
of these contributions seem to favor the $N_c$-counting scheme; meson-loop 
contributions have been found to be suppressed. Arguments as to why this is the 
case have been presented in Ref.~\cite{Melnikov:2003xd}.

Strictly speaking, however, one does not actually perform a large-$N_c$
expansion as this would necessitate the inclusion of an infinite number
of resonances. Instead, only the lowest lying meson exchange contributions 
in the pseudoscalar, scalar and axialvector channel have been
subsummed.
Here, the pseudoscalar $\pi^0$-exchange has been identified as the leading
contribution, followed by $\eta$ and $\eta'$-exchange.

Concerning the pseudoscalar (PS) exchange contribution a few remarks are in place.
The photons in the exchange diagrams are coupled to the PS mesons
via the PS-$\gamma\gamma$ form-factor, $F_{\mathrm{PS}\gamma\gamma}$.
It is evident that there are two limiting features of the pseudoscalar-pole
approximation. The first is the actual provision of the form-factors themselves,
which are in general subject to systematic errors depending on how they are
modeled or calculated. The second is the procedure under which the
form-factor is taken off-shell. Previous approaches mainly used vector meson
dominance ideas to determine this form factor and there has been an
extensive debate as to whether and how short distance constraints have to
be implemented 
\cite{Knecht:2001qf,Melnikov:2003xd,Dorokhov:2008pw,Nyffeler:2009tw,Prades:2009tw}. 
Rather than employ the principles of vector-meson dominance and construct an 
ansatz for the on-shell/off-shell form-factor, we wish to calculate it from
first principles. This is possible within the framework of Dyson-Schwinger
and Bethe-Salpeter equations using a well explored and successful truncation 
scheme \cite{Maris:2003vk}.

As for the quark-loop diagram, different interpretations have been given in the
literature. Whereas in \cite{deRafael:1993za} it has been argued that the quark
loop is a separate contribution that has to be added to the other two, in many
other approaches it has been treated as a complementary one, which is only added
in the large spacelike momentum region, say above a typical cutoff for an effective 
model. In our functional approach, described below, it is clearly the first point
of view that is correct. Moreover, as we will see, the quark loop is subject to large
dressing effects not only for the quark propagators in the loop but also for the
quark-photon vertices. This will be the main result of our work.

\subsection{Expansion using functional methods}

From a functional integral approach to QCD, featuring quarks and gluons
as the fundamental degrees of freedom, the analogous picture to what is
normally considered in the literature is shown in
Fig.~\ref{fig:photon4pt_2}, where we give an expansion in terms of 
non-perturbatively dressed one-particle irreducible Green's functions.
The basic idea of this expansion is {\it not} a separation of long distance 
and short distance scales, but rather a separation of different classes 
of diagrams based on their topology. Clearly, the expansion is such that 
no double counting of diagrams is
involved. By considering a restricted subset of
contributions in which only diagrams with a planar topology are
resummed, we effectively adhere to the $N_c$-counting scheme as favored
in the EFT approaches mentioned above. Though there are similarities 
between the two pictures, since we work
with a truncated formulation of exact-QCD rather than an effective field
theory there are some differences that we will comment on here to avoid
confusion. First of all, our quarks are to be interpreted in the same
way as those extracted via Lattice QCD; they are characterised by
momentum dependent dressing functions that interpolate between the
current and constituent quark limits, {\it cf.} the discussion below
Fig.~\ref{fig:mtquarks}. Secondly, the quark-photon
coupling is a non-perturbative form-factor and not merely a tree-level
bare vertex; it can be calculated self-consistently for a given
truncation scheme. Finally, the planar resummation of gluons is related
to the $T$-matrix of quark-antiquark scattering and contains meson poles
that can be associated with pseudoscalars, vectors, scalars etc. This will
be exploited below, where we return to the conventional meson exchange
picture to approximate these contributions.

We wish to emphasize that the expansion displayed in Fig.~\ref{fig:photon4pt_2}
has been used successfully in a different context already in Ref.~\cite{Bicudo:2001jq}. 
There $\pi-\pi$ scattering has been considered using similar quark-box and 
ladder exchange parts as displayed in Fig.~\ref{fig:photon4pt_2}. In this 
setup, the authors of Ref.~\cite{Bicudo:2001jq} could reproduce the isospin 0 and 2 
scattering lengths in exact agreement with Weinberg's low energy results. 
Moreover, in Ref.~\cite{Cotanch:2002vj} it has been checked, that the corresponding 
resonant expansion similar to the one displayed in Fig.~\ref{fig:photon4pt} 
is a good approximation to the ladder exchange part of Fig.~\ref{fig:photon4pt_2}.
Note that in both these calculations the quark-box diagram had to be added
to the ladder exchange or the resonant 'meson-exchange' part respectively. 
We believe that these results add further support to our approach.

\subsubsection{Quark-loop contribution\label{sec:quarkloop}}
\begin{figure}[t]
\begin{equation*}
\begin{array}{c}
\includegraphics[height=2.2cm]{photon4ptfn-qrkloop}
\end{array}
\!\!=\!\!
\begin{array}{c}
\includegraphics[height=2.2cm]{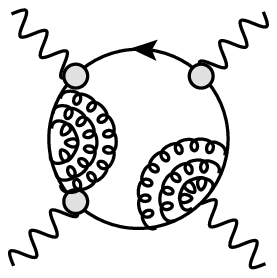}
\end{array}
\!\!+\!\!
\begin{array}{c}
\includegraphics[height=2.2cm]{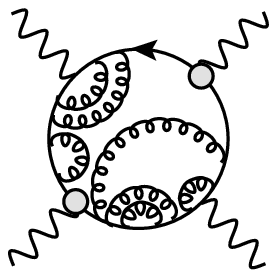}
\end{array}
\!\!+\cdots
\end{equation*}
\caption{Expansion of quark-loop contribution to the photon four-point function
in terms of planar quark and gluon diagrams (all propagators are fully dressed).
\label{fig:planar}}
            \includegraphics[width=0.3\columnwidth]{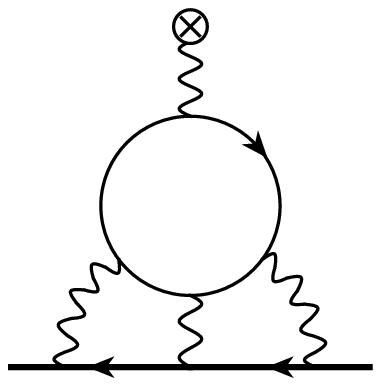}
            \includegraphics[width=0.3\columnwidth]{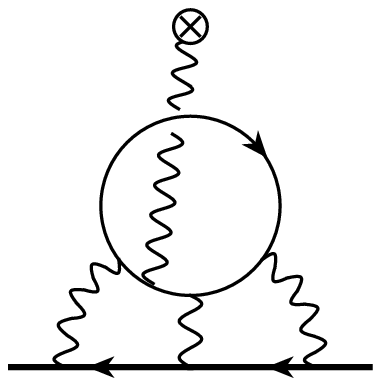}
            \includegraphics[width=0.3\columnwidth]{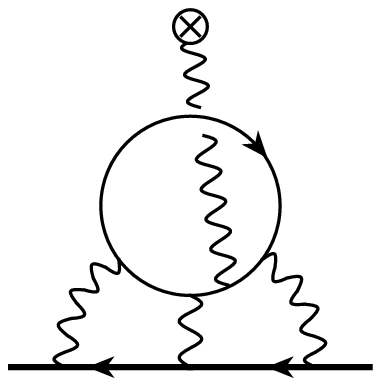}
      \caption{The hadronic light-by-light scattering contributions to
               $a_\mu$ from the quark loop. There are an additional
		   three diagrams (not shown) in which the quark spin-line
		   is reversed. Principally, these diagrams involve dressed
		   quark propagators and quark-photon vertices.\label{fig:quarkloop}}
      \end{figure}

Within our proposed truncation, the quark-loop is composed of dressed
quark propagators and dressed quark-photon vertices. On expanding these
one-particle irreducible Green's functions, within the rainbow-ladder
approximation, we find planar-like diagrams such as the ones
shown in Fig.~\ref{fig:planar} (all propagators 
are fully dressed),
where in fact infinite ladders of gluons are taken into account. Should
we consider corrections beyond rainbow-ladder, such as those considered
in Ref.~\cite{Fischer:2007ze,Fischer:2008wy}, one would also include
diagrams in which the gluons have self-interactions as well as crossed-ladder
components. Taking into account such corrections is, however, beyond the
scope of the present work.

Considering this contribution to the muon-photon vertex, we obtain the 
diagrams as shown in Fig.~\ref{fig:quarkloop}, where we have shown
permutation of the external photon legs but have omitted the topologies 
that merely involve reversal of the quark-spin line (these give identical
contributions and hence constitute a factor of two). As is well-known, these diagrams are 
individually logarithmically divergent with only their sum finite and convergent; 
thus one employs the aforementioned trick, Eq.~(\ref{eqn:DefOfFivePointFunction}), 
of taking the derivative of the photon
four-point function~\cite{Aldins:1969jz,Aldins:1970id}

Since this is now to be applied to loop integrals over non-perturbative
quantities, namely the quark-propagator, it is no longer possible to 
reduce the integration to be five-dimensional as in the case of 
perturbative studies. More generally, on considering the planar 
nature of the diagrams, one must deal with 8-dimensional integrals which
necessitate Monte-Carlo methods~\cite{Hahn:2004fe}.
However, for reasons of calculational simplicity we actually integrate 
in nine. We did check, however, that we were able to reproduce the well-known
perturbative results for the electron loop contribution to the anomalous
magnetic of the electron and the muon~\cite{Kinoshita:1988yp,Samuel:1992ub,Laporta:1992pa} . Additionally, 
due to the somewhat involved Dirac algebra~\cite{Vermaseren:2000nd,Reiter:2009ts} we will content ourselves with
taking the quark-photon vertices inside the quark-loop contribution to 
be: (a) bare; (b) 1BC; (c) full BC. The precise meaning of these abbreviations 
and the relation to the full quark-photon vertices will become clear in 
sections \ref{sec:quarkphotonvertex} and \ref{res:quarkloop}. 
The extension to employ the numerically 
calculated non-perturbative form of the vertex will be explored in 
a later publication. The results of our calculation are presented in
section~\ref{sec:results}.

\subsubsection{Ladder-exchange and ladder-ring contribution}
\begin{figure}[t]
\begin{equation}
\begin{array}{c}
\includegraphics[height=2.2cm]{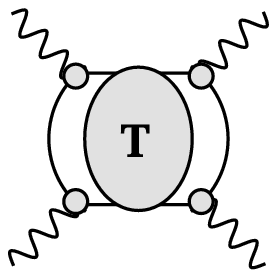}
\end{array}
=
\begin{array}{c}
\includegraphics[height=2.2cm]{photon4ptfn-ladder}
\end{array}
+\cdots
\end{equation}

\begin{equation}
\begin{array}{c}
\includegraphics[height=2.2cm]{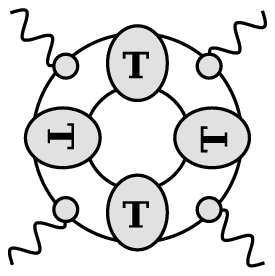}
\end{array}
=
\begin{array}{c}
\includegraphics[height=2.2cm]{photon4ptfn-ring}
\end{array}
+\cdots
\end{equation}
\caption{Ladder-exchange contribution (upper equation) and 
ring-ladder contribution (lower equation) to the photon four-point amplitude.\label{fig:ladder_ring}}
	\centering
\begin{eqnarray*}
\begin{array}{c}
\includegraphics[scale=0.8]{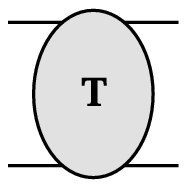}
\end{array}
&=&
\begin{array}{c}
\includegraphics[scale=0.8]{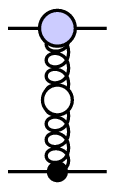}
\end{array}
\!+\!
\begin{array}{c}
\includegraphics[scale=0.8]{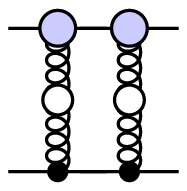}
\end{array}
\!+\!
\begin{array}{c}
\includegraphics[scale=0.8]{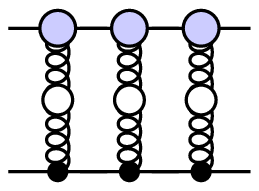}
\end{array}
+\cdots
\\
\begin{array}{c}
\includegraphics[scale=0.8]{tmatrix-amp}
\end{array}
&=&
\begin{array}{c}
\includegraphics[scale=0.8]{tmatrix-1l}
\end{array}
\!+\!
\begin{array}{c}
\includegraphics[scale=0.8]{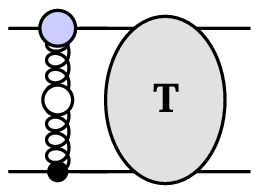}
\end{array}
\\
\begin{array}{c}
\includegraphics[scale=0.8]{tmatrix-amp}
\end{array}
&&\xrightarrow{P^2\rightarrow-M^2}
\begin{array}{c}
\includegraphics[scale=0.8]{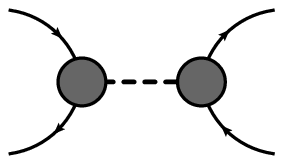}
\end{array}
\end{eqnarray*}
\caption{The $T$-matrix in Rainbow-Ladder approximation. (a) shows the
series expansion in terms of dressed quarks and gluon, whilst (b)
represents Dyson's equation. (c) Shows the pole-ansatz for the T-matrix
on-mass shell.\label{fig:tmatrix}}
\end{figure}

Two contributions that are leading and sub-leading in large-$N_c$
respectively are the so-called ladder-exchange and ladder-ring diagrams
of Fig.~\ref{fig:photon4pt_2}. These infinite ladder resummations are in
fact related to the $T$-matrix of bound-state theory in a certain
approximation scheme (that produces planar diagrams). Thus, another way
to portray these contributions is given in Fig.~\ref{fig:ladder_ring}.
The $T$-matrix in rainbow-ladder approximation is given in
Fig.~\ref{fig:tmatrix}. 
At this point, we make it clear that there is no conflict nor double
counting between the quark-loop and ladder-exchange diagrams, as they
clearly consider and resum different topologies of diagrams.

\begin{figure}[t]
\begin{eqnarray*}
\begin{array}{c}
\includegraphics[height=2.0cm]{photon4ptfn-ladder}
\end{array}
\xrightarrow{P^2\rightarrow-M_{\mathrm{PS}}^2}
\begin{array}{c}
\includegraphics[height=2.0cm]{photon4ptfn-pole}
\end{array}
\end{eqnarray*}
\caption{Pole representation of the ladder-exchange contribution to the
photon four-point function. \label{fig:polepart}}
            \includegraphics[width=0.40\columnwidth]{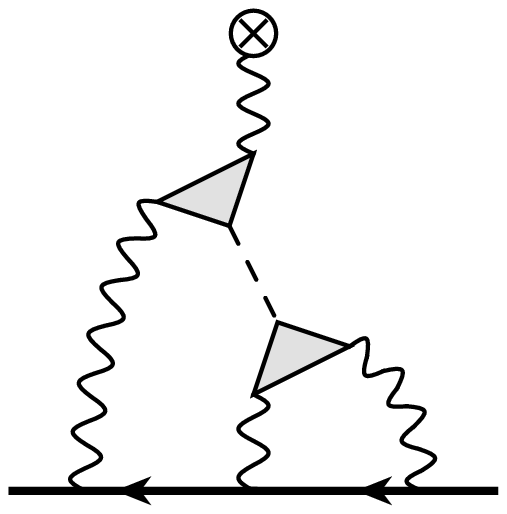}
      \caption{The pion-pole part of the LBL contribution to $a_\mu$. 
               The three possible permutations of the photon legs are not shown.
               \label{fig:lblpionpole}}
      \end{figure}

As it stands, the full $T$-matrix is a very complicated object
to solve in its entirety though its structure admits several
approximations and simplifications~\cite{Watson:2004jq}. 
The one which we employ here is
similar to the viewpoint taken by Effective Field Theory approaches;
that is, we consider pole contributions to be dominant. Now, since it is
well-known that such an \emph{infinite} gluon-ladder resummation
dynamically generates bound-state poles, one can expand the $T$-matrix
in terms of meson pole contributions as shown in
Fig.~\ref{fig:tmatrix}(c). On mass shell we then have a unique
definition of the Bethe-Salpeter amplitude, described below in
Fig.~\ref{fig:bse}, that gives the form-factor describing
coupling of a meson to two quarks. From this point the
(on-shell) pseudoscalar-photon-photon form-factor can be defined and 
calculated, giving rise to the `leading' pseudoscalar meson exchange part,
as shown in Fig.~\ref{fig:polepart} and Fig.~\ref{fig:lblpionpole}.

In a similar fashion, the ring-ladder diagram contains contributions
akin to the pion loop on meson mass-shell. However, since these are
generally considered to be sub-leading we will not consider them further
here and instead concentrate on the quark-loop and ladder-exchange
diagrams. After we present our approach and formalism in the next section,
results will be discussed in section \ref{sec:results}.

\section{Framework\label{sec:model}}
\begin{figure}[t]
      \includegraphics[width=0.97\columnwidth]{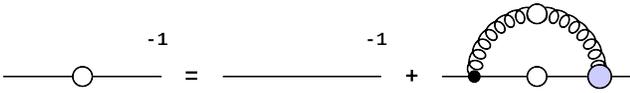}
      \caption{Dyson--Schwinger equation for the quark propagator. Specification
               of the fully-dressed gluon propagator (wiggley line) and 
               quark-gluon vertex (coloured blob) defines the truncation 
               scheme.}\label{fig:quarkdse}
\end{figure}

The dressed quark propagator is one of the most important quantities in the covariant 
description of mesons. It encodes non-perturbative properties of QCD
such as dynamical 
mass generation and the realization of a non-zero condensate. Its equation of
motion, the quark Dyson-Schwinger equation (DSE) displayed in Fig.~\ref{fig:quarkdse}, 
also contains the dressed gluon propagator and a dressed quark-gluon vertex. 
Whereas the dressed gluon propagator in Landau gauge is a well-known quantity by now,
see~\cite{Bonnet:2001uh,Fischer:2003rp,Kamleh:2007ud,Fischer:2008uz} and references 
therein\footnote{There is an intense debate on the behaviour
of the gluon propagator in the deep infrared, i.e. for momenta $p \le 50$ MeV.
It seems, however, that this momentum region is irrelevant when it comes
to the calculation of observables \cite{Fischer:2007ze,Fischer:2008wy,Fischer:2009gk}.}, the
study of the details of the dressed quark-gluon vertex is still on an exploratory
level, although some progress has been made in the past years 
\cite{Fischer:2008wy,Skullerud:2003qu,Alkofer:2008tt}. Pending deeper insights
into the nonperturbative structure of the quark-gluon interaction it is therefore 
reasonable to work with approximations that take into account important features
of the full theory. This strategy, of course, introduces model dependencies into
our calculation that have to be carefully addressed later on.

>From a phenomenological perspective a successful approximation in this respect is the
rainbow-ladder truncation of the quark-DSE. The philosophy here is to combine the dressing 
of the gluon propagator with the vector part of the quark-gluon vertex into a single function
depending on the gluon momentum only. While this is certainly a severe approximation in
principle, in practice it turned out to be very successful as concerns the calculation of 
masses and electromagnetic properties of mesonic observables \cite{Maris:2003vk}. 
While the parameters of the model are tuned such that it reproduces the experimental values 
for the masses and decay constants of the pion, it also reproduces the pion charge radius 
and $\pi \gamma \gamma$ transition form factors on the percent level. In the vector channel 
the agreement with experimental masses and decay constants is on the five and ten percent 
level. Thus, while one has to keep in mind possible systematic caveats, we nevertheless
believe that such a model is an excellent starting point for a systematic evaluation of
hadronic LBL.

In Euclidean momentum space, the renormalized dressed gluon and 
quark propagators in the Landau gauge are given by
	\begin{align}
  		D_{\mu\nu}(p) &=  \left( \delta_{\mu\nu}-\frac{p_\mu p_\nu}{p^2}  
  		\right)\frac{Z(p^2;\mu^2)}{p^2}           \,,\label{eqn:gluon}\\[2mm]
  		S_F(p) &=  \frac{Z_f(p^2;\mu^2)}{i\sh{p}  + M(p^2)} =
  		\frac{1}{i\sh{p} A(p^2;\mu^2)+B(p^2;\mu^2)}\,,  \label{eqn:quark}
	\end{align}
where $Z(p^2;\mu^2)$ is the gluon dressing function, $Z_f(p^2;\mu^2)$ is the
quark wave-function and $M(p^2)$ is the renormalisation point independent 
quark mass function. The dependence of such functions on the renormalisation 
point $\mu^2$ will be implicitly assumed from here on. The quark dressing 
functions $A(p^2)$ and $B(p^2)$ can be recombined into the quark mass and 
wave-function by $M(p^2) = B(p^2)/A(p^2)$ and $Z_f(p^2) = 1/A(p^2)$.

These propagators may be obtained by solving their respective
Dyson--Schwinger equations. The DSE for the quark propagator, shown
diagrammatically in Fig.~\ref{fig:quarkdse}, is written
	\begin{align}
		S^{-1}(p) &= Z_{2}S^{-1}_{0}(p)
		+\Sigma(p)\;,\notag\\[-2mm]\label{eqn:quarkdse}\\[-2mm]\notag
		\Sigma(p) &= g^{2}C_{F}Z_{1F}\int\frac{d^4q}{\left( 2\pi \right)^4}
		\Gamma_{\nu}(q,p)  D_{\mu\nu}(k) 
		\gamma_{\mu}S_F(q)\;,
	\end{align}
where $\Sigma(p)$ is the quark self-energy, $k=p-q$ and the Casimir $C_F=4/3$ 
stems from the colour trace. We
introduced the reduced quark-gluon vertex $\Gamma_\nu(q,p)$
defined by $\Gamma_\nu^a(q,p)=ig \frac{\lambda^a}{2}\Gamma_\nu(q,p)$. The bare
inverse quark propagator is $S^{-1}_{0}(p) = i \sh{p} + m$. The
renormalisation
factors are $Z_{1F}=Z_2/\widetilde{Z}_3$ for the quark-gluon vertex, $Z_2$
for the quark propagator and $\widetilde{Z}_3$ for the ghost dressing function.

\begin{figure}[t!]
      \includegraphics[width=0.9\columnwidth]{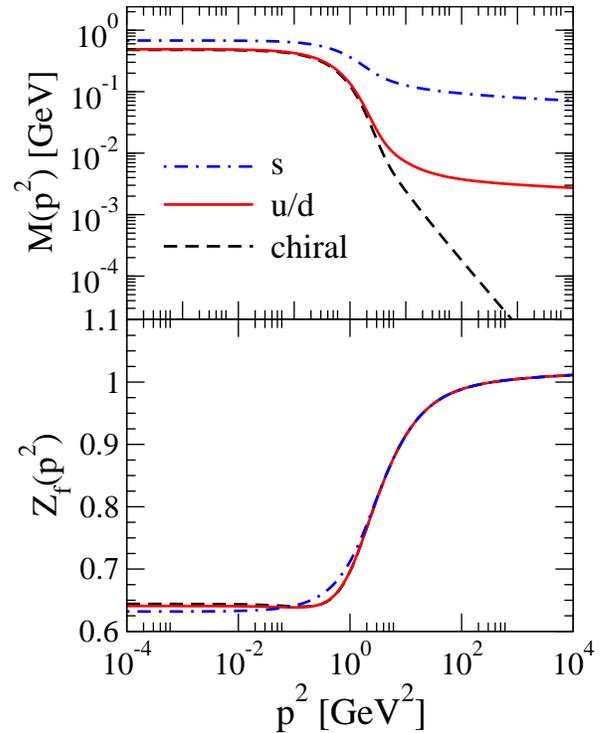}
      \caption{Mass function and wave-function dressing functions
	corresponding to quark propagators solved with the Maris-Tandy
	interaction \cite{Maris:1999nt}.}\label{fig:mtquarks}
\end{figure}

The scalar dressing functions of the quark DSE are solved for by
appropriate projections of Eq.~(\ref{eqn:quarkdse}). This is a
coupled non-linear integral equation that is solvable provided we know
the gluon dressing function and the structure of the quark-gluon vertex.
In the rainbow approximation both are specified by Ans\"atze, with in particular
the choice $\Gamma_\nu(q,p):=\Gamma^{\mathrm{YM}}(k^2)\gamma_\nu$, with
scalar function $\Gamma^{\mathrm{YM}}$ representing the non-perturbative
dressing of the quark-gluon vertex and $k=q-p$ the gluon momenta.
Here, the gluon dressing function $Z(k^2)$ from Eq.~(\ref{eqn:gluon})
and the Yang-Mills part $\Gamma^{\rm YM}(k^2)$ of the quark-gluon vertex 
are combined to form a phenomenological effective interaction.
For the Maris-Tandy (MT) model~\cite{Maris:1999nt} this function 
is given by
	\begin{eqnarray}
	Z(k^2) \Gamma^{\mathrm{YM}}(k^2) &=&  \frac{4\pi}{g^2} 
      	 \bigg( \frac{\pi}{\omega^6}D k^4 \exp(-k^2/\omega^2)\nonumber\\
		&+&\frac{2\pi \gamma_m}{\log\left( \tau+\left(1+k^2/\lqcdsq \right)^2\right)}
		\nonumber\\[0.mm]
		&&\times \left[ 1-\exp\left(-k^2/\left[ 4m_t^2 \right]\right) \right]\bigg)\;, 
      	\label{eqn:maristandy}
	\end{eqnarray}
with
	\begin{equation}
		\begin{array}{lcl}
		m_t= 0.5\;{\rm GeV}\,&,&\qquad \tau\;=\;\mathrm{e}^2-1\,\\
 		\gamma_m=12/(33-2N_f)\,&,&\quad \lqcd\;=\;0.234\,{\rm GeV}\, .
 		\end{array}\nonumber
	\end{equation}
This interaction corresponds to a Gaussian distribution in the infrared that 
provides for sufficient interaction strength to generate DCSB, together with 
the one-loop behavior of the running coupling at large, perturbative, momenta. 
The latter is mandatory to provide for the correct short distance behavior
of the quark propagator. The remaining parameters $\omega$ and $D$ essentially 
constitute a single one-parameter family of solutions for which pion observables 
remain comparable, via $\omega D=(0.72 \mbox{GeV})^3$. 

For the convenience of the reader, in Fig.~\ref{fig:mtquarks} we again 
show the two dressing functions $Z_f(p^2)$ and $M(p^2)$ that characterize 
the non-perturbative quark propagator, obtained by solving 
Eq.~(\ref{eqn:quarkdse}) using the Maris-Tandy interaction \cite{Maris:1999nt}. 
Clearly, in the mass function there are three distinguished momentum regions. In the infrared, the
quark propagator is essentially constant displaying the behavior of a constituent 
quark. Then for $1~\mbox{GeV}^2 < p^2 < 10~\mbox{GeV}^2$ there is a region of 
rapid change, where the quark mass function follows the well known $1/p^2$ behavior 
expected from the operator product expansion. For even larger momenta and
non-vanishing current quark mass the quark mass function behaves logarithmically
as expected for a current quark. The fully dressed quark propagator thus naturally 
interpolates between the constituent and current quark picture. 
We consider this feature of the Dyson-Schwinger approach to QCD as an advantage 
compared with effective models such as the ENJL model.

\subsection{Bethe-Salpeter equation\label{sec:bse}}
\begin{figure}[b!]
\begin{eqnarray*}
\begin{array}{c}
\includegraphics[scale=1.0]{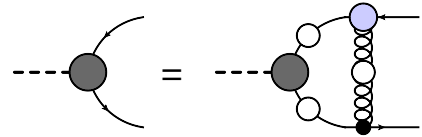}
\end{array}
\end{eqnarray*}
\caption{The homogeneous Bethe-Salpeter equation for the meson
amplitude.\label{fig:bse}}
\end{figure}

The chiral symmetry preserving truncation for the Bethe-Salpeter equation, 
consistent with the rainbow-approximation above, is given by the 
ladder approximation
\begin{widetext}
	\begin{align}\label{eqn:bse}
		\Gamma_{tu}^{q\bar{q}}(p;P)&= \int \frac{d^4k}{(2\pi)^4}K_{tu;rs}(p,k;P) 
			   \left[S_F(k_+)\Gamma^{q\bar{q}}(k;P)S_F(k_-)\right]_{sr}
	\end{align}
with the kernel $K_{tu;rs}$ given by
	\begin{align}  
 		K_{tu;sr}(q,p;P)
  		&=
  		\frac{g^2 \, Z(k^2)\,  \Gamma^{\textrm{YM}}(k^2) \, Z_{1F}}{k^{2}}
  		\left(\delta_{\mu\nu}-\frac{k_{\mu}k_{\nu}}{k^{2}}\right)
  		\left[\frac{\lambda^{a}}{2}\gamma_{\mu}\right]_{ts}
  		\left[\frac{\lambda^{a}}{2}\gamma_{\nu}\right]_{ru}\,, \label{eqn:kernel}
	\end{align}
\end{widetext}
see Fig.~\ref{fig:bse} for a graphical representation.
Here $\Gamma^{q\bar{q}}(p;P)$ is the Bethe-Salpeter vertex function 
corresponding to a pseudoscalar quark anti-quark bound-state, specified below.
The momenta
$k_+=k+P/2$ and $k_-=k-P/2$ are such that the total momentum $P$ 
of the meson is given by $P=k_+-k_-$ and the relative momentum
$k=(k_++k_-)/2$. The Latin indices ($t,u,r,s$) of the
kernels refer to colour, flavour and Dirac structure. 

The form of the kernel 
Eq.~(\ref{eqn:kernel}) is uniquely determined from the axial-vector Ward-Takahashi
identity and ensures, that the pion is a Goldstone-boson in the chiral limit
without any fine-tuning of parameters. It also ensures that important constraints 
from chiral symmetry such as the Gell-Mann-Oakes-Renner relation are satisfied.

In general, the covariant structure of the Bethe-Salpeter vertex function, 
$\Gamma^{q\bar{q}}(p;P)$,
determines the quantum numbers of the bound-state under consideration. In
particular, a pseudoscalar meson is completely specified by the following form
	\begin{align}
  		\Gamma^{q\bar{q}}(p;P)&= \gamma_{5}\Big[F^{q\bar{q}}_1(p;P)
               -i\Sh{P} F^{q\bar{q}}_2(p;P)\notag
            \\[-3mm]\label{eqn:pion}\\[-3mm]
               &-i\sh{p} \left(p\cdot P\right)F^{q\bar{q}}_3(p;P)
                -\left[\Sh{P},\sh{p}\right]F^{q\bar{q}}_4(p;P)\Big]
            \;\; .\notag
	\end{align}
This amplitude is obtained through solution of \Eq{eqn:bse} on-mass shell: 
$P^2=-m_{q\bar{q}}^2$ in Euclidean space. While (\ref{eqn:pion})
represents a quark-anti-quark bound-state, physical mesons are
defined as matrices in flavour space built out of the $q\bar{q}$ amplitudes.
This then leads to the same decomposition as in \Eq{eqn:pion} but with 
flavour matrix valued quantities $\Gamma$ and $F_i$. In the following, however,
we will keep the flavour index implicit and use \Eq{eqn:pion} for $q\bar{q}$ amplitudes
and mesons alike. Anyway, in the isospin-limit considered herein the pion amplitude 
differs from the $u\bar{u}/d\bar{d}$ amplitudes only by flavour-matrix structure. 
The pole masses and the scalar amplitudes that contain the dynamical information 
are identical (up to normalization). This is different for the $\eta$ and $\eta'$.

\begin{figure}[t]
      \includegraphics[width=0.9\columnwidth]{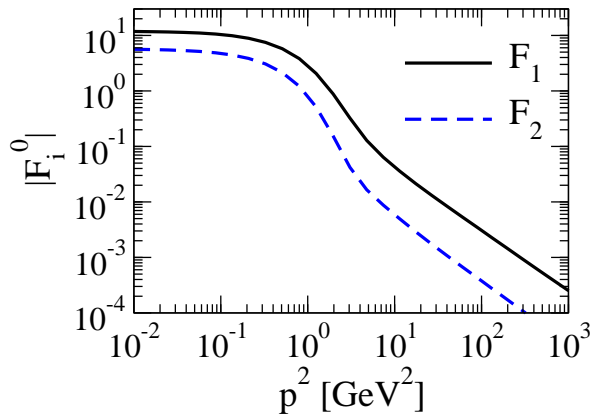}
      \caption{Normalised Bethe-Salpeter amplitudes $F_1$ and $F_2$ of the 
      pion solved with the Maris-Tandy interaction (leading
	Chebyshev component)\cite{Maris:1999nt}.}\label{fig:pion}
\end{figure}

In the chiral limit the leading behavior of the pion amplitude $\hat{\Gamma}^{\pi}$ 
is given by
      \begin{align}\label{eqn:chiralpion}
	\hat{F}_1^{\pi}(p;P) :=\lambda_3 B(p^2)/f_\pi
      \end{align}
where $B(p^2)$ is the scalar dressing function of the quark,
$f_\pi$ is the chiral limit value of the leptonic decay constant, and $\lambda_3$ 
is a Gell-Mann matrix that represents the flavour structure. The hat in \Eq{eqn:chiralpion} 
indicates that the object is matrix valued in flavour space. The expressions for 
the calculation of $f_\pi$ and the normalization condition of the
Bethe-Salpeter amplitude together with details on the numerical procedure
for dealing with the BSE are given in \cite{Maris:1997hd}. 
For the convenience of the reader we display the resulting Bethe-Salpeter amplitudes
for $p^2=-m_\pi^2$ and $p.P=0$ in Fig.~\ref{fig:pion}.
Qualitatively, the amplitudes have a similar form as the quark mass function
in the chiral limit. For large momenta and up to logarithmic corrections they 
fall off like $1/p^2$, which is a necessary condition to correctly describe the 
anomalous decay of the pion and to reproduce the asymptotics of the pion form 
factor \cite{Maris:1998hc}.

\subsection{Quark-photon Vertex\label{sec:quarkphotonvertex}}

\begin{figure}[b]
      \includegraphics[width=0.95\columnwidth]{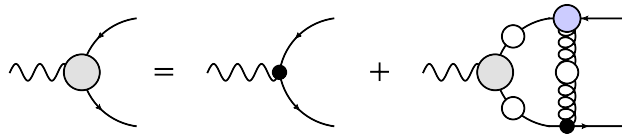}
      \caption{Inhomogeneous BS equation for the quark-photon vertex, in
               rainbow-ladder approximation.}\label{fig:quarkphotondse}
\end{figure}
An important quantity for the determination of the LBL contribution to
the muon $g-2$
is the dressed quark-photon vertex. This quantity is genuinely non-perturbative
in nature and necessary for the calculation of the $\textrm{PS}\gamma\gamma$ form-factor. It
describes the coupling of a fully dressed quark to a photon and is dominated
by QCD corrections. As a function of one Lorentz and two spinor indices, it can
be decomposed into twelve Dirac structures
	\begin{align}
		\Gamma_\mu(P,k) = \sum_{i=1}^{12} \lambda_i(P,k) V^{i}_\mu(P,k) \label{vertex}
	\end{align}
where $V^{i}_\mu(P,k)$ represents the basis components, and $\lambda_i(P,k)$ the
non-perturbative dressing functions. A common basis is that of Ball and 
Chiu \cite{Ball:1980ay}, in which $V^{i}_\mu(P,k)$ is split into
terms that are transverse and non-transverse with respect to the photon momentum.
The Ward-Takahashi identity and regularity assumptions constrain the form of the
non-transverse part in terms of quark propagator functions
	\begin{align}    \label{eqn:ballchiuansatz}
		\Gamma^{\mathrm{BC}}_\mu(k,P):=&\Big[ \gamma_\mu \frac{A(k_+^2)+A(k_-^2)}{2}\notag\\
       	+& (\sh{k}_++\sh{k}_-)(k_++k_-)_\mu\frac{1}{2}\frac{A(k_+^2)-A(k_-^2)}{k_+^2-k_-^2}\notag\\
       	+&i(k_++k_-)_\mu\frac{B(k_+^2)-B(k_-^2)}{k_-^2-k_+^2}\Big],
	\end{align}
leaving only the strictly transverse pieces undetermined. \Eq{eqn:ballchiuansatz} can
therefore be seen as an approximation of the full vertex and has been used in situations,
where the full vertex cannot be determined numerically.

A more sophisticated approach, however, is to solve the inhomogeneous Bethe-Salpeter equation
of the quark-photon vertex, shown diagrammatically in 
Fig.~\ref{fig:quarkphotondse}. It is given by
	\begin{align}
  	&\Gamma^\mu_{tu}(k,P)= Z_{1}\gamma^\mu_{tu} +
	\int_q K_{tu;rs}\big[ S_F(k_+)\Gamma^\mu(q,P)S_F(k_-)\big]_{sr}
  	\label{eqn:quarkphotondse}
	\end{align}
where $Z_1$ is the renormalisation factor associated with the quark-photon
vertex. By using the same interaction kernel as in the BSE for mesons, \Eq{eqn:kernel}, 
not only do we achieve self-consistency within the truncation scheme, but also 
by virtue of its symmetry preserving nature we satisfy the Ward-Takahashi identity. 
Consequently, the non-transverse part of the vertex, given in \Eq{eqn:ballchiuansatz}, is 
nicely reproduced numerically, with transverse terms additionally generated 
\cite{Maris:1999bh,Maris:1999ta,Maris:2002mz,Bhagwat:2006pu}.

We note here that such a determination of the quark-photon vertex automatically
contains poles in the time-like region corresponding to vector meson exchange.
Thus, presupposing that vector-meson dominance is an important feature in the
structure of the pion electromagnetic form-factor, it is already included here
as a result of the approach we employ. This has been discussed in detail also in 
Refs.~\cite{Maris:1999bh,Maris:1999ta}.

The numerical details involved in the calculation of the quark-photon vertex
have been described in several works, see e.g. the appendix of Ref.~\cite{Bhagwat:2006pu}.
Below we will use the fully dynamical, selfconsistent solution of \Eq{eqn:quarkphotondse}
for our calculation of the $\textrm{PS}\rightarrow\gamma\gamma$ form-factor and the resulting 
meson exchange contribution to LBL. Unfortunately, because of the numerical 
complexity we have to restrict ourselves to the exact longitudinal part given by 
\Eq{eqn:ballchiuansatz} in the quark-loop diagram. It will be a subject of future
work to overcome this limitation.

\subsection{The $\textrm{PS}\rightarrow\gamma\gamma$ form-factor\label{sec:formfactor}}
\begin{figure}[t]
      \includegraphics[width=0.4\columnwidth]{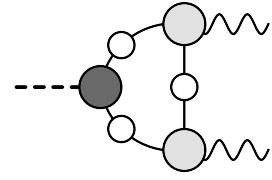}
      \caption{The $\pi^0\gamma\gamma$ form-factor in impulse approximation. All
               internal quantities are fully-dressed.}\label{fig:quarktriangle}
\end{figure}

The coupling of the exchanged pseudoscalar mesons to photons is the quantity that
is central to the resonant expansion of Fig.~\ref{fig:photon4pt}. In impulse
approximation, consistent with the rainbow-ladder truncation scheme introduced
in sections \ref{sec:model}, the Bethe-Salpeter amplitude
is connected via a quark-triangle to the fully-dressed quark-photon vertex,
as shown in Fig.~\ref{fig:quarktriangle}. For a pseudoscalar $\mathrm{PS}$ we have
	\begin{align}
            \Lambda_{\mu\nu}^{\mathrm{PS}\gamma^*\gamma^*}(k_1,k_2) &=\!   
	    2e^2N_c\!\!\int_k \tr\big[ i\hat{\mathcal{Q}}_e\Gamma_\nu(k_2,p_{12})S_F(p_2)\notag\\
	    \;\;\;\;\times
	    \hat{\Gamma}^{\mathrm{PS}}(p_{23}&,P)S_F(p_3)i\hat{\mathcal{Q}}_e\Gamma_\mu(k_1,p_{31})S_F(p_3)\big]\;,
		\label{eqn:PseudoScalarFormFactor}
	\end{align}
where $k_1$ and $k_2$ are the outgoing photon momenta, $p_1=q$, $p_2=q-k_2$ and $p_3=q+k_1$ are the quark momenta and 
$p_{ij}=\left(p_i+p_j\right)/2$. The factor of two stems from exchange of the two photon vertices
and $\hat{\mathcal{Q}}=\mathrm{diag}\left[ 2/3,-1/3,-1/3 \right]$ gives
the quark's charge. The PS vertex
$\hat{\Gamma}^{\mathrm{PS}}$ is explicitly matrix valued in flavour space. It is defined as
	\begin{align}
		\pi^0:& \qquad \hat{\Gamma}^{\pi^0} = \frac{1}{\sqrt{2}}
		\mathrm{diag}\left[\Gamma^{u\bar{u}},-\Gamma^{d\bar{d}},0\right]\notag\\
		\eta^8:& \qquad \hat{\Gamma}^{\eta^8} =  \frac{1}{\sqrt{6}}
		\mathrm{diag}\left[\Gamma^{u\bar{u}},\Gamma^{d\bar{d}},-2\Gamma^{s\bar{s}}\right]
		\label{eqn:flavourStructure}\\
		\eta^0:& \qquad \hat{\Gamma}^{\eta^0} =  \frac{1}{\sqrt{3}}
		\mathrm{diag}\left[\Gamma^{u\bar{u}},\Gamma^{d\bar{d}},\Gamma^{s\bar{s}}\right]\notag,
	\end{align}
for the pseudoscalar mesons. The $\Gamma^{q\bar{q}}$ are solutions of Eq.~(\ref{eqn:bse}).
In addition we work in the isospin-limit ($\Gamma^{u\bar{u}} =\Gamma^{d\bar{d}}$).
Since the quantities in \Eq{eqn:flavourStructure} are defined in the singlet-octet basis
we have to rotate in order to obtain the $\eta$-$\eta'$ amplitudes
\begin{align}
  	\hat{\Gamma}^\eta   \, &= \cos\theta\,\hat{\Gamma}^{\eta^8} - \sin\theta\,\hat{\Gamma}^{\eta^0}\notag 
			 \\[-3mm]\label{eqn:EtaEtaPrimeRotation}\\[-3mm]\notag
	\hat{\Gamma}^{\eta'}   &= \sin\theta\,\hat{\Gamma}^{\eta^8} + \cos\theta\,\hat{\Gamma}^{\eta^0}\notag,
\end{align}
where we haven taken $\theta = -15.4^{\,\circ}$ \cite{Feldmann:1998vh}.
The pseudoscalar electromagnetic form-factor can be described by a single scalar 
function, $F^{\pi\gamma^*\gamma^*}$. For the pion this function can be given a natural 
normalization via the Abelian anomaly~\cite{WZWABJ:Anomaly}
      \begin{align}\label{eqn:scalarformfactor}
           \Lambda^{\pi\gamma^*\gamma^*}_{\mu\nu}(k_1^2,k_2^2) &= 
            i \frac{\alpha_{\mathrm{em}}}{\pi f_\pi} \varepsilon_{\mu\nu\alpha\beta}
            k_1^{\alpha}k_2^{\beta} F^{\pi\gamma^*\gamma^*}(k_1^2,k_2^2)\;\;,
      \end{align}
where $\alpha_{\mathrm{em}}$ is the fine structure constant and $f_\pi$ the pion decay constant.
The definition of the prefactors is such that $F^{\pi\gamma\gamma}(0,0)=1$. The $\eta$- and $\eta'$-mesons have the same tensor structure. 

Note that the form factors determined here do not accurately reflect all effects due to the
topological mass of the $\eta_0$, simply because the $U_A(1)$-anomaly is not represented
correctly in the Maris-Tandy model\footnote{Perspectives to improve this issue in the framework of
Dyson-Schwinger equations have been reported in 
Refs.~\cite{vonSmekal:1997dq,Bhagwat:2007ha,Alkofer:2008et}. In Ref.~\cite{Alkofer:2008et} a
topological mass of the $\eta_0$ has been obtained which goes well with lattice results
of the topological susceptibility via the Witten-Veneziano relation.}. 
In the form factors this may be a minor
problem. The effect is larger, however, in the meson propagators attached to the
form factors. We therefore prefer to use the experimental masses in these propagators 
thereby taking care of the majority of the $U_A(1)$-anomaly effects.

\begin{figure}[t]
      \includegraphics[width=0.95\columnwidth]{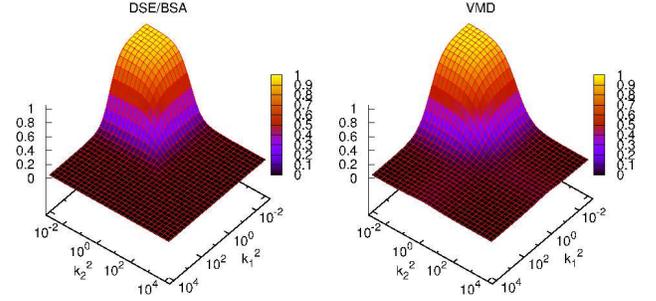}
      \caption{The $\pi\rightarrow\gamma\gamma$ form-factor plotted as a function
      of the two photon momenta $k_1^2$ and $k_2^2$. We compare our numerical results 
      in the Dyson-Schwinger/Bethe-Salpeter approach with an ansatz inspired by
      vector meson dominance discussed in Ref.~\cite{Nyffeler:2009tw}. }\label{fig:FF}
\end{figure}

The $\pi^0$ electromagnetic form-factor has been explored in detail in Ref.
\cite{Maris:2002mz}, wherein it has been confirmed that
the correct normalization is satisfied. In addition it has been shown analytically 
(and numerically) that the correct asymptotic behaviour, modulo
potential logarithms is obtained~\cite{Roberts:1999,Maris:2002mz},
\begin{align}
\lim_{Q^2\rightarrow\infty}F^{\pi^0\gamma\gamma^*}(0,Q^2)\propto&\frac{1}{Q^2}\;\;
\label{asymmetricLimit}\nonumber\\
	\lim_{Q^2\rightarrow\infty}F^{\pi^0\gamma^*\gamma^*}(Q^2,Q^2)\propto&\frac{1}{Q^2}\;\;.
\end{align}
In Fig.\ref{fig:FF} we plot the form factor as a function of the two photon momenta 
$k_1^2$ and $k_2^2$ and compare with the VMD inspired model used in 
Ref.~\cite{Knecht:2001qf}.
One clearly sees that both form factors agree nicely on a qualitative and even quantitative 
level. Whereas the low-momentum behaviour is governed by the anomaly, at large momenta
both form factors fall off according to Eqs.~(\ref{asymmetricLimit}). There are small 
quantitative differences in the mid-momentum regime, which will lead to a small difference
in the meson exchange contributions to LBL, discussed below. In general, however, the
results of our calculation may be viewed as a confirmation of the previously used model
approaches almost from first principles. 

\subsection{Off-shell prescription}

It is evident from the kinematics of the diagram  shown in Fig.~\ref{fig:lblpionpole} that the form-factors,
thus far defined as on-shell quantities, must be evaluated for momenta of the
exchanged pseudoscalar meson that would be far from the pole mass. In the
approach considered here, the pseudoscalar amplitude is obtained from its
homogeneous Bethe-Salpeter equation and hence is by definition an on-shell 
quantity. Thus, to proceed we must introduce a prescription for the continuation
of this quantity to the off-shell momentum region.

Since the off-shell behaviour should be dominated by the pseudoscalar
pole contribution, the introduction of any prescription that provides
for a suppression at off-shell momentum should be suitable starting point. 
Here, we will employ a prescription that is inspired
from the axial-vector Ward-Takahashi identity in the chiral limit
\begin{align}\label{eqn:chiralaxwti}
  2P_\mu \Gamma^{5,3}_\mu(k,P) = i S^{-1}(k_+)\gamma_5 + i\gamma_5 S^{-1}(k_-)\;\;.
\end{align}
Here the axial vector vertex is defined as the correlation function
$\Gamma^{5,3}_\mu=\langle j_{\mu}^{5,3}q\bar{q}\rangle$ that includes the axial vector current
in the pion channel $j_{\mu}^{5,3}= \bar{q}\gamma_\mu\gamma^5\frac{\lambda^3}{2}q$. 
It is clear that Eq. (\ref{eqn:chiralaxwti}) relates $\Gamma^{5,3}_\mu$ to the
quark propagator. Taking explicit parameterisations for vertex and propagator 
(see \cite{Maris:1997hd})
that include the pion pole in the axial vector vertex,
the following form of the dominant amplitude for the $\pi^0$ can be deduced:
\begin{align}\label{eqn:offshellprescriptionScalar}
  \hat{F}^{\pi}_1(k,P)=\lambda_3\frac{B(k_+)+B(k_-)}{2f_\pi}\;.
\end{align}
Here $k_{\pm} = k\pm P/2$. Note that in the chiral on-shell limit ($P^2=0$)
the above equation reduces to Eq.~(\ref{eqn:chiralpion}).
We generalize the pseudoscalar amplitude Eq.~(\ref{eqn:pion}) by  using
Eq.~(\ref{eqn:offshellprescriptionScalar}) 
as a guideline for all four  structures also away from the chiral limit. The final off-shell meson  amplitude reads:
\begin{align}
  \hat{\Gamma}^{\mathrm{PS}} =&  \gamma_{5}\Big[\hat{F}^{\mathrm{PS}}_1(p;P)
	  +f(P^2)\Big\{-i\Sh{P} \hat{F}^{\mathrm{PS}}_2(p;P)\notag
                          \\[-3mm]\label{eqn:pionOffShell}\\[-3mm]
			  &      -i\sh{p} \left(p\cdot P\right)\hat{F}^{\mathrm{PS}}_3(p;P)
			  -\left[\Sh{P},\sh{p}\right]\hat{F}^{\mathrm{PS}}_4(p;P)\Big\}\Big],\notag
\end{align}
where the hat over the functions $\hat{F}^{\mathrm{PS}}_i$ indicates that the flavour
structure of the corresponding meson is included in the same manner as
in Eqs.~(\ref{eqn:flavourStructure},~\ref{eqn:EtaEtaPrimeRotation}). 
The scalar off-shell amplitudes 
$F^{q\bar{q}}_i(k,P)$ are defined in terms of the on-shell amplitudes
\footnote{With a slight abuse of notation we denote off-shell quantities 
to depend on $P$ whereas on-shell they depend only on $k\cdot P$ with 
$P^2=-m^2_{\textrm{PS}}$.}
$F^{q\bar{q}}_i(k,k\cdot P)$ 
through
\begin{align}\label{eqn:offshellprescription}
	F^{q\bar{q}}_i(k,P)=\frac{F^{q\bar{q}}_i(k_+,k_+\cdot P)+F^{q\bar{q}}_i(k_-,k_-\cdot P)}{2}\;,
\end{align}
for which $i=1,\ldots,4$. The on-shell amplitudes are obtained via Eq.~(\ref{eqn:bse}).
In order to account for the mass dimensions of the form-factors $\hat{F}^{\mathrm{PS}}_{2,3,4}$ we attach
to each the function
\begin{align}
  f(P^2) = \sqrt{\frac{m^2_{\mathrm{PS}}}{P^2 + 2 m^2_{\mathrm{PS}}}}\;\;.
  \label{eqn:supressionFactor}
\end{align}
This prevents an unnatural enhancement at high meson virtuality whilst
at the same time leaving the on-shell behaviour unchanged.
The off-shell form factor
\begin{align}
	\Lambda_{\mu\nu}^{\mathrm{PS}^*\gamma^*\gamma^*}(P,k_1,k_2) = &
	\epsilon_{\mu\nu\alpha\beta}k_1^\alpha k_2^\beta
	\F_{\mathrm{PS}^*\gamma^*\gamma^*}(P^2,k^2_1,k^2_2)\;\;,
\end{align}
is then obtained via the generalisation of Eq.~(\ref{eqn:PseudoScalarFormFactor})
by taking the Bethe-Salpeter amplitude $\Gamma^{q\bar{q}}$ to be defined via 
Eq.~(\ref{eqn:pionOffShell}). 

The contribution to the derivative of the four-point-function can now be written as \cite{Knecht:2001qf}:
\begin{widetext}
\begin{align}
	\widetilde{\Pi}_{(\rho)\mu\nu\lambda\sigma}(q_1,q_2,-q_{12}) 
	&=
	\frac{\F_{\mathrm{PS}^*\gamma^*\gamma^*}(q_{12}^2,q_1^2,q_2^2)
  		\F_{\mathrm{PS}^*\gamma^*\gamma^*}(q_{12}^2,q_{12}^2,0) }{q_{12}^2+m^2_{\mathrm{PS}}}
  		\epsilon_{\mu\nu\alpha\beta}q_1^{\alpha}q_2^\beta \epsilon_{\lambda\sigma\rho\tau}q_{12}^\tau \notag\\
      &+
	 \frac{\F_{\mathrm{PS}^*\gamma^*\gamma^*}(q_1^2,q_1^2,0)
  		\F_{\mathrm{PS}^*\gamma^*\gamma^*}(q_1^2,q_2^2,q_{12}^2) }{q_1^2+m^2_{\mathrm{PS}}}
  		\epsilon_{\mu\sigma\tau\rho}q_1^{\tau} \epsilon_{\nu\lambda\alpha\beta}q_1^{\alpha}q_2^\beta  \notag\\ 
	&+
	\frac{\F_{\mathrm{PS}^*\gamma^*\gamma^*}(q_2^2,q_1^2,q_{12}^2) 
  		\F_{\mathrm{PS}^*\gamma^*\gamma^*}(q_2^2,q_2^2,0)}{q_2^2+m^2_{\mathrm{PS}}}
  		\epsilon_{\mu\lambda\alpha\beta}q_1^{\alpha}q_2^\beta
  		\epsilon_{\nu\sigma\rho\tau}q_2^\tau\;\;,
    \label{eqn:Resonant4ptFunction}
\end{align}
\end{widetext}
where $q_{12}=q_1+q_2$. The function $\widetilde{\Pi}_{(\rho)\mu\nu\lambda\sigma}$ is now only dependent upon two momenta
since the limit $k\rightarrow 0$ for the external photon momentum has 
been carried out.

\section{Results\label{sec:results}}
With the approach to hadronic LBL scattering within the functional
approach outlined, and our truncation scheme defined we proceed to
combine our propagators, vertices, amplitudes and form-factors together
and calculate the respective contributions to hadronic LBL scattering in
the muon $g-2$.

\begin{table*}[t!]
\renewcommand{\tabcolsep}{1.5pc}	
\renewcommand{\arraystretch}{1.2}	
\begin{tabular}{@{}l|l|cc}
Group                               			& 
Model                               			& 
$a_\mu^{\mathrm{LBL}}$~($\pi^0$-pole)			&
$a_\mu^{\mathrm{LBL}}$~(quark-loop) 			\\
\hline
\hline
Bijnens,~Prades,~Pallante  \cite{Bijnens:1995cc}      &
ENJL                                			&
$59(11)$                          				&
\hspace*{-2.5mm}$21(3)$                          	\\
Hayakawa, Kinoshita~\cite{Hayakawa:1997rq}, HK and Sanda \cite{Hayakawa:1995ps,Hayakawa:1996ki}         &
HLS                                 			&
\hspace*{-1.5mm}$57(4)$                          	&
\hspace*{2.5mm}$9.7(11.1)$                            \\
Knecht and Nyffeler  \cite{Knecht:2001qf}         	&
LMD+V                               			&
$58(10)$                          				&
--									\\
Melnikov and Vainshtein  \cite{Melnikov:2003xd}       &
LMD+V                               			&
\hspace*{-1.5mm}$77(5)$                          	&
$\dag$								\\
Dorokhov and Broniowski  \cite{Dorokhov:2008pw}       &
NL$\chi$QM    							&
\hspace*{-1.5mm}$65(2)$                      		&
$\ddag$								\\
Nyffeler \cite{Nyffeler:2009tw}           		&
LMD+V                               			&
$72(12)$                          				&
$\dag$								\\
\hline
Our Result			                          	&
DSE                                         		&
$58(10)$                            			&
\hspace*{-2.5mm}$136(59)$					\\
\end{tabular}

\caption{Results for the $\pi^0$-pole and quark-loop contribution (where
appropriate) to hadronic light-by-light scattering, in different models.
For $\dag$ the quark-loop correction is incorporated as a boundary
condition on the pion-pole contribution, whilst for $\ddag$ the quark-loop
corrections are currently under investigation~\cite{Dorokhov:prep}.\label{tab:results}}
\end{table*}

\subsection{Pion-pole contribution to LBL}
To demonstrate parity between our approach and others in the
determination of hadronic LBL, we calculate the ladder-exchange diagram
of Fig.~\ref{fig:photon4pt_2} assuming pseudoscalar pole-dominance. Once
more, we re-iterate that on mass-shell this is \emph{identical} to the
pseudoscalar exchange diagram portrayed in Fig.~\ref{fig:photon4pt},
whereas off-shell we make the common assumption that the meson-exchange picture 
provides a good approximation.

In order to determine the pseudoscalar exchange contribution we must 
numerically determine the dressed quark propagator,
the quark-photon vertex, and the homogeneous Bethe-Salpeter amplitude
for the pseudoscalar meson. Combining these together allows us to calculate
from first principles the $\pi\gamma\gamma$ form-factor. We wish to 
emphasize again that the resulting quark-photon vertex 
also contains time-like poles corresponding to vector meson 
exchange~\cite{Maris:1999ta}. Thus the main ideas of VMD are naturally 
included here in the form-factor due to the non-perturbative approach 
that we employ. We have checked that the total numerical error of our 
calculation is of the order of one percent. In a similar fashion we also 
evaluate the corresponding form factors for the $\eta$ and $\eta'$ mesons.
We then use our results for the form factors to evaluate the pseudoscalar 
meson exchange contribution to LBL. For this, we use the off-shell
prescription for the pseudoscalar Bethe-Salpeter amplitude proposed in
Eq.~(\ref{eqn:pionOffShell}) for the exchanged mesons. This prescription
gives a reduction of the contribution that is similar to that found in
other approaches. 

The systematic error of our calculation of the pseudoscalar exchange diagrams
can be attributed entirely to the validity of the rainbow-ladder
approximation, by the Maris-Tandy (MT) model, \Eq{eqn:maristandy}, and the off-shell prescription 
for the mesons, \Eq{eqn:offshellprescriptionScalar}. No other approximations have 
been used. While in the Goldstone-Boson sector the MT model works well, there 
is certainly a larger error in the flavour singlet sector. We therefore
guesstimate a total systematic error: ten percent for the pion contribution, 
and twenty percent for the $\eta$ and $\eta'$ contributions. With a numerical 
error of two percent we then obtain:
$a_\mu^{\textrm{LBL};\pi^0}=(57.5 \pm 6.9)\times 10^{-11}$,
$a_\mu^{\textrm{LBL};\eta} =(13.6 \pm 3.0)\times 10^{-11}$ and
$a_\mu^{\textrm{LBL};\eta'}=( 9.6 \pm 2.1)\times 10^{-11}$ leading to
	\begin{align}
		a_\mu^{\textrm{LBL;PS}}=(80.7 \pm 12.0)\times 10^{-11} \label{res:PS}
	\end{align}
for the pseudoscalar meson exchange contribution to LBL. As compared to our
previous work \cite{Fischer:2010iz}, the values for $a_\mu^{\textrm{LBL};\eta}$ 
and $a_\mu^{\textrm{LBL};\eta'}$ are slightly reduced due to a more consistent
off-shell prescription in these channels. Our result (\ref{res:PS}) is
compatible with previous ones \cite{Nyffeler:2009tw,Prades:2009tw,Dorokhov:2004ze},
which for the pion pole contribution are displayed in Table~\ref{tab:results}.
This is not surprising, since the form-factors themselves are compatible
at a qualitative level, {\it cf}. Fig.~\ref{fig:FF}, and all approaches make 
the assumption that pseudoscalar pole-dominance is valid far from the meson 
mass-shell.\\

\subsection{Quark loop contribution to LBL \label{res:quarkloop}}
Having convinced ourselves that the method works we now focus on the quark loop
contribution to LBL. Here we follow the strategy described around \Eq{eqn:DefOfFivePointFunction}
where the Ward identity obeyed by the four-point function with respect to the external 
field is exploited to construct quantities that are explicitly finite. As for our 
numerical error we verified that we reproduce the well known perturbative result 
for the corresponding electron loop with an accuracy of better than one per mille. 
In the quark loop we use the fully dressed quark propagators for the up, down, 
strange and charm quarks, extracted from the DSE, Fig.~\ref{fig:quarkdse}. 
As already mentioned above, due to numerical complexity we are unfortunately not 
yet in a position where we can use the full, numerically determined quark-photon 
vertex used for the meson exchange contributions in the previous section. Instead,
we use three different approximations to the full vertex and compare the results.
As explained in section \ref{sec:quarkphotonvertex} due to the Ward-identity we
are in possession of exact expressions for the non-transversal, Ball-Chiu part of 
the vertex, \Eq{eqn:ballchiuansatz}. We exploit this knowledge to compare results
with (a) a bare vertex, (b) the first term of \Eq{eqn:ballchiuansatz} (1BC)
	\begin{align}
		\Gamma_\mu(p,q) = \frac{A(p^2)+A(q^2)}{2} \,\,\gamma_\mu \; \label{vertexdressing}
	\end{align}
where $p,q$ are the quark and antiquark momenta, and (c) the full Ball-Chiu (BC) expression
\Eq{eqn:ballchiuansatz}. A comparison between these three approximations may serve as 
a guide for the systematic error due to the relevance of vertex effects. We emphasize,
however, that only our most elaborate approximation, (c), satisfies the
constraints of gauge invariance.
Previous approximations based on purely transverse parts of the vertex \cite{Bijnens:1995cc} 
do not satisfy this constraint. We believe that the ansatz (c) provides an excellent 
basis for the calculation of the quark-loop diagram, which can and should be expanded
in future work to also include transverse parts of the vertex.

As a result of our calculation we find
	\begin{equation}
		\begin{array}{lcc}
		a_\mu^{\textrm{LBL;quarkloop (bare vertex)}} &=&
		(\phantom{0}61 \pm 2) \times 10^{-11}\\
		a_\mu^{\textrm{LBL;quarkloop (1BC)}}         &=& 
		(107 \pm 2)		    \times 10^{-11}\\
		a_\mu^{\textrm{LBL;quarkloop (BC)}}         &=& 
		(176 \pm 4)		    \times 10^{-11}\\
		\end{array}\label{res:QL} \\
		\end{equation}
for the quark loop contribution. Clearly these are sizable contributions. Whereas the
bare vertex result roughly agrees with the number $60 \times 10^{-11}$ given in 
\cite{Melnikov:2003xd}, the dressing effects of the vertex lead to a drastic 
increase. As compared to our result for the first part of the Ball-Chiu vertex 
\cite{Fischer:2010iz} we again find a drastic increase from $107 \times 10^{-11}$
to $176 \times 10^{-11}$ when the other two terms of the full Ball-Chiu vertex
are included. In this calculation we included effects from four quark flavours
in the quark-loop. Their individual contributions are given by 
	\begin{equation}
		\begin{array}{lcc}
		a_\mu^{\textrm{LBL;quarkloop (BC);u/d}} &=&
		(158 \pm 3) \times 10^{-11}\\
		a_\mu^{\textrm{LBL;quarkloop (BC);s}}         &=& 
		(\phantom{0}6 \pm 1)		    \times 10^{-11}\\
		a_\mu^{\textrm{LBL;quarkloop (BC);c}}         &=& 
		(\phantom{0}12 \pm 1)		    \times 10^{-11}\\
		\end{array}\label{res:QLudsc} \\
	\end{equation}
It is interesting to note that due to charge effects the heavy charm 
quark contributes more than the much lighter strange quark. 

We have checked the model dependence of the above result by comparing with a 
similar calculation using a different model for the quark-gluon interaction 
\cite{Fischer:2003rp}. The results are similar to the one in \Eq{res:QL} 
within an error margin of five to ten percent. Details will be given elsewhere. 
Due to these results we estimate an additional systematic 
error for our BC-result of $15 \times 10^{-11}$, which has to be added to the 
$4 \times 10^{-11}$ given in \Eq{res:QL}.

In general, these large dressing effects also make it very hard if not impossible
to guess the effect of the total vertex dressing without an explicit calculation. 
Certainly, however, given these findings, all previous estimates for the 
systematic error in the quark loop contributions seem to be an order of 
magnitude too small.

\section{Conclusions\label{sec:conclusion}}
In this paper we have presented a new approach towards the anomalous 
magnetic moment of the muon. We have used a combination of Dyson-Schwinger 
and Bethe-Salpeter equations to evaluate the pseudoscalar meson exchange 
contribution and the quark loop contribution to LBL. Our only input is 
the Maris-Tandy model, a phenomenologically successful ansatz for the 
combined strength of the gluon propagator and the quark-gluon vertex.
Our treatment of the meson exchange contribution to LBL is different from
earlier approaches in that we do not rely on an ansatz for the $\textrm{PS} \gamma \gamma$
form factor, but calculate this quantity starting from the basic equations
of motion of QCD. Nevertheless, our result basically agrees with those from 
previous approaches. This result once more emphasizes that the meson exchange
contributions to LBL are largely controlled by analytic constraints from QCD
at large and small $Q^2$.

As for the quark-loop contribution, analytic constraints have been used which 
arise from the requirement of gauge invariance: the quark-photon vertex appearing 
in this loop has to satisfy the vector Ward-Takahashy identity. In contrast to
previous approaches, we have implemented this identity by using the 
Ball-Chiu ansatz for this vertex. We believe this is a systematic improvement. 
The consequences are drastic: we observe a dramatic increase for the quark-loop 
contribution to LBL. Our result of $(176 \pm 4) \times 10^{-11}$ is more than 
three times larger than the constituent quark result of Ref.~\cite{Melnikov:2003xd}.

When combining our two results, \Eq{res:PS} and \Eq{res:QL}, we arrive at a 
hadronic LBL contribution of 
	\begin{equation}
		\begin{array}{lcc}
			a_\mu^{\textrm{LBL;PS+quarkloop}} &=& (257 \pm 31)\times 10^{-11}\;. \\
		\end{array}
	\end{equation}
This value, however, does not yet account for transverse parts of the quark-photon
vertex in the quark-loop contribution and for effects from the right hand diagrams 
of Fig.~\ref{fig:photon4pt} or Fig.~\ref{fig:photon4pt_2}. In general, it is 
difficult to gauge the effects of additional transverse vertex contributions in the 
quark-loop diagram. In Ref.~\cite{Bijnens:1995cc} part of these effects have been taken
into account by using an ansatz motivated by vector-meson dominance (VMD) ideas. 
They found a reduction due to these effects of roughly $40 \times 10^{-11}$. Since 
we agree with Ref.~\cite{Bijnens:1995cc} on the size of the pion exchange contribution,
where VMD works very well, it may be justified to use their result as a rough estimate
for the size and also for the potential error in these effects. We therefore add
a contribution\footnote{Note that cross-terms between transverse and non-transverse contributions from the four vertices in the quark-box diagram could provide additional
suppression. This is expressed in our error estimate.} 
of $a_\mu^{\textrm{LBL;quarkloop,transverse}} = (-40 \pm 40)\times 10^{-11}$
to arrive at $a_\mu^{\textrm{LBL;quarkloop (BC+transverse)}} = (136 \pm 59)\times 10^{-11}$.

The additional contributions due to the right hand diagrams 
of Fig.~\ref{fig:photon4pt} or Fig.~\ref{fig:photon4pt_2} are also difficult to judge.
It may help, though, to observe that these involve an additional quark-loop. Typically
such contributions are negative and of the order of ten to twenty percent of the leading-$N_c$ 
contributions \cite{Fischer:2007ze,Fischer:2008wy}. Since on the other hand one also 
expects positive contributions of a similar size from non-pseudoscalar exchange 
diagrams \cite{Jegerlehner:2009ry} we choose to subsume all these contributions to 
another $a_\mu^{\textrm{LBL;other}}=(0 \pm 20)\times 10^{-11}$, where the error is 
clearly subjective. This gives us the following total hadronic LBL
contribution 
	\begin{equation}
		\begin{array}{lcc}
		a_\mu^{\textrm{LBL}} &=& (217 \pm 91)\times 10^{-11}\;, \\
		\end{array}
	\end{equation}
in our approach. Note that the increase of the central value as compared to our previous
result in Ref.~\cite{Fischer:2010iz} is due to a combination of taking the full BC-vertex 
instead of 1BC and in addition accounting for the transverse corrections in the
quark-loop using the results of Ref.~\cite{Bijnens:1995cc}.
Taken at face value these numbers together with the other contributions
quoted in \cite{Jegerlehner:2009ry} clearly reduce the discrepancy between theory and 
experiment. Combining our light-by-light scattering results with the
other SM contributions gives:
	\begin{align}\label{eqn:newamu}
		a_\mu^{\textrm{theor.}}=116\,591\,891.0(105.0)\times 10^{-11}\;\;.
	\end{align}

To put this result in perspective we wish to recall the caveats that to our 
mind are tied to it. First, there is the contribution of transverse parts of 
the quark-photon vertex to the quark-loop diagram. Although the results of 
Ref.~\cite{Bijnens:1995cc} may serve as an estimate, we definitely need to 
explicitly calculate these contributions in our approach. Second, there is
the question whether the pseudoscalar meson exchange diagram provides for
a good approximation of the gluon exchange contribution discussed around 
Fig.~\ref{fig:photon4pt_2}. Also this assumption needs to be questioned by
an explicit calculation. In this sense, our results certainly do not provide 
final answers but still have to be seen as a further step towards a fundamental 
determination of $a_\mu$. 

Finally, we point out that the current approach will also be checked by a 
calculation of the hadronic vacuum polarization contribution
to $a_\mu$. Preliminary results in this direction are encouraging.

\begin{acknowledgments}
This work was supported by the Helmholtz-University Young Investigator Grant 
No.~VH-NG-332 and by the Helmholtz International Center for FAIR within the 
LOEWE program of the State of Hesse.
\end{acknowledgments}

\end{document}